 \documentclass[10pt,conference]{IEEEtran} 
\IEEEoverridecommandlockouts
\usepackage{cite}
\usepackage{amsmath,amssymb,amsfonts}
\usepackage{algorithmic}
\usepackage{graphicx}
\usepackage{textcomp}
\usepackage{xcolor}

\usepackage{multirow}
\usepackage{subfigure}
\usepackage{graphicx}
\usepackage{enumitem}
\usepackage{graphicx}
\usepackage{balance}
\usepackage[ruled,linesnumbered]{algorithm2e}
\usepackage{float}
\usepackage{booktabs}
\usepackage{flushend}
\usepackage{color}
\usepackage{soul}
\usepackage{url}
\usepackage{makecell}

\newtheorem{definition}{Definition}[section]
\newtheorem{assumption}{Assumption}[section]

\def\BibTeX{{\rm B\kern-.05em{\sc i\kern-.025em b}\kern-.08em
    T\kern-.1667em\lower.7ex\hbox{E}\kern-.125emX}}
\begin{document}

\newboolean{showcomments}
\setboolean{showcomments}{true} 
\ifthenelse{\boolean{showcomments}}
{\newcommand{\nb}[2]{
  \fcolorbox{black}{yellow}{\bfseries\sffamily\scriptsize#1}
  {\sf\small$\blacktriangleright$\textit{#2}$\blacktriangleleft$}
 }
 \newcommand{\version}{\emph{\scriptsize$-$working$-$}}
}
{\newcommand{\nb}[2]{}
 \newcommand{\version}{}
}
\newcommand\Haiyan[1]{\nb{Haiyan}{#1}}

\newboolean{showcomments-2}
\setboolean{showcomments-2}{false} 
\ifthenelse{\boolean{showcomments-2}}
{\newcommand{\nbz}[2]{
  \fcolorbox{black}{yellow}{\bfseries\sffamily\scriptsize#1}
  {\sf\small$\blacktriangleright$\textit{#2}$\blacktriangleleft$}
 }
 \newcommand{\versionz}{\emph{\scriptsize$-$working$-$}}
}
{\newcommand{\nbz}[2]{}
 \newcommand{\versionz}{}
}
\newcommand\zhi[1]{\nbz{Zhi}{#1}}


\title{A Meta Reinforcement Learning-based Approach for Self-Adaptive System}
\author{\IEEEauthorblockN{Mingyue Zhang}
\IEEEauthorblockA{\textit{Peking University}, Beijing, China\\ mingyuezhang@pku.edu.cn}
\\
\IEEEauthorblockN{Kenji Tei}
\IEEEauthorblockA{\textit{Waseda University}, Tokyo, Japan\\ ktei@aoni.waseda.jp}
\and
\IEEEauthorblockN{Jialong Li}
\IEEEauthorblockA{\textit{Waseda University}, Tokyo, Japan\\ lijialong@fuji.waseda.jp}
\\
\IEEEauthorblockN{ Shinichi Honiden}
\IEEEauthorblockA{\textit{Waseda University}, Tokyo, Japan\\ honiden@nii.ac.jp}
\and 
\IEEEauthorblockN{Haiyan Zhao}
\IEEEauthorblockA{\textit{Peking University}, Beijing, China\\ zhhy.sei@pku.edu.cn}
\\
\IEEEauthorblockN{Zhi Jin}
\IEEEauthorblockA{\textit{Peking University}, Beijing, China \\ zhijin@pku.edu.cn}
}

\maketitle

\begin{abstract}
A self-learning adaptive system (SLAS) uses machine learning to enable and enhance its adaptability. 
Such systems are expected to perform well in dynamic situations. 
For learning high-performance adaptation policy, some assumptions must be made on the environment-system dynamics when information about the real situation is incomplete. 
However, these assumptions cannot be expected to be always correct, and yet it is difficult to enumerate all possible assumptions. 
This leads to the problem of incomplete-information learning. 
We consider this problem as \emph{multiple model problem} in terms of finding the adaptation policy that can cope with multiple models of environment-system dynamics. 
This paper proposes a novel approach to engineering the online adaptation of SLAS. 
It separates three concerns that are related to the adaptation policy and presents the modeling and synthesis process, with the goal of achieving higher model construction efficiency.
In addition, it designs a meta-reinforcement learning algorithm for learning the meta policy over the multiple models, so that the meta policy can quickly adapts to the real environment-system dynamics. 
At last, it reports the case study on a robotic system to evaluate the adaptability of the approach.

\end{abstract}
\section{Introduction}
Adaptation policy is key to \emph{self-adaptive system}, as it determines the behavior that the system needs to take in response to the environment-system dynamics \footnote{Here, the \emph{environment} refers to the physical spatial environment, and the \emph{system} refers to the targeted system. In many cases, the environment-system dynamics can be modeled as automata.} and/or the changes of users' goals.
That is even more prominent in \emph{self-adaptive space systems} \cite{tsigkanos2016Architecting}.
In self-adaptive systems, adaptation policy is normally constructed as a function, which maps the states of a system to its actions \cite{sutton1998introduction, ma2012rule, lanese2010framework}. 
For enhancing the system's adaptability, \emph{self-learning adaptive systems} (SLAS) are proposed to uses machine learning techniques to learn the adaptation policy.
This topic has attracted a lot of attentions from academia and industry \cite{elkhodary2010fusion, jamshidi2017transfer, knauss2016acon} and is more important for \emph{self-adaptive space systems} because modern cyber-physical spaces are much more dynamic than traditional self-adaptive systems.


Generally speaking, an SLAS is expected to learn from its interactions with the environment to generate and evolve its adaptation policy with environment dynamics \cite{kim2009reinforcement, zhao2017reinforcement, ho2015model}.
One of the difficulties that hinder the SLAS from learning high-performance adaptation policy is the problem of \emph{incomplete information}, i.e. some crucial information about the environment-system dynamics may be missing during learning.
For precisely modeling the the environment-system dynamics, it is suggested to give some assumptions that may result in different dynamics models \cite{daniel2020multitier, yang2016selfadaptation, zhang2020survey}, so that, the SLAS can use these models to learn different adaptation policies.


The \emph{multiple model problem}, i.e. to find an adaptation policy that can cope with multiple environment-system dynamics models, needs to be taken into account when designing an SLAS.
The literature considers this as a problem of \emph{assumption violation}. 
Inevitably, the assumptions about the environment-system dynamics may not be the same with the actual dynamics.
The policy learned based on the assumptions should be repaired through the exploration of different models to suit the actual dynamics\cite{daniel2020multitier, nicol2014hope}. 
Most of the existing studies deal with the multiple model problem by taking a ``Re-'' approach, which can be of two categories. 
One is \emph{revising the model}; that is, the discrepancies between the offline assumptions and real environment-system dynamics are reflected in the models \cite{sykes2013learning, nicol2014hope, daniel2020multitier, jamshidi2017transfer, nair2018transfer}. 
When current assumptions are violated, the model is revised and based on the revised models, an adaptation policy is re-planned for run-time usage. 
The other is \emph{directly relearning a policy}; that is, a new adaptation policy is learned when discrepancies exist between the model and reality. 
The re-learning methods include re-learning from scratch \cite{kim2009reinforcement, tesauro2007reinforcement} and re-learning from an existing policy whose model is most similar to the actual dynamics \cite{zhao2017reinforcement, elkhodary2010fusion}.

However, the ``Re-'' approaches, especially ones based on machine learning, are time and resource consuming, resulting in performance degradation at run-time \cite{kim2009reinforcement, javier2016analyzing}. 
Moreover, in \emph{self-adaptive space systems}, another challenge is the increasing shift of merely static physical spaces to highly dynamic cyber-physical ones.
When different aspects related to the adaptation policy are intertwined, it is inevitably difficult to explore, compare and learn the environment-system dynamics model. 
Considering the assumptions from separate or different viewpoints based on the principle of \emph {Separation of Concerns} (SoC) will ease the task of model revision and therefore will help the policy learning in SLAS.

This paper focuses on the self-adaptive space systems and proposes a novel approach, named \emph{\underline{me}ta \underline{r}einforcement learning \underline{a}daptive \underline{p}lanning} (MeRAP).
This approach enables the SLAS to quickly adjust its policies online at the cost of offline training. 
The main ideas are: 
(1) constructing separate models capturing different concerns in self-adaptive space systems, and then semi-automatically integrating them into multiple models represented by Markov decision processes (MDPs);
(2) using the meta reinforcement learning algorithm to allow the SLAS to learn from multiple models and enable the system dynamically and quickly adjust its adaptation policies according to the actual environment-system dynamics. 

For point (1), we model the environment-system dynamics based on the principle of \emph{separation of concerns} (SoC). 
We distinguish three viewpoints \cite{tsigkanos2018early,nianyu2020early} that are related to the adaptation policy, i.e., the spatial environment, the system capabilities, and the objectives.
Correspondingly, we separate the dynamics model into three kinds of  sub-models, i.e., the environment models, the capability models, and the objective models. 
The integrated dynamic models are obtained by synthesizing these sub-models.

For point (2), we design a meta reinforcement learning (MRL) algorithm to train a meta policy over these integrated models.
The objective of the MRL algorithm is to minimize the sum of losses for all candidate models so that the meta policy learned by MRL can fast adapt to any model among these candidates. 
Further, if there is a sufficient number of candidate models, the meta policy can even fast adapt to models that have never been seen before.

To evaluate the effectiveness of this approach, we conduct a case study on a robotic system, and make the comparison with two baseline approaches, i.e. the online policy evolution \cite{kim2009reinforcement, zhao2017reinforcement} and the directly planning through a pre-trained policy. 
The experimental results show that the re-planning time of our approach is significantly reduced to 0.63\% of that of online policy evolution, and learns the optimal adaptation policy in most cases.


The main contributions are threefold:
\begin{itemize}
\item 
A novel approach, i.e. MeRAP, is proposed which integrates meta reinforcement learning into MAPE-K (i.e. monitoring, analysis, planning, execution and knowledge) self-adaptive loop.
 
\item 
A process is devised to model different design concerns separately for easing the model revision and to support the model synthesis for enabling meta reinforcement learning.
 
\item 
A meta reinforcement learning for a meta policy over the multiple models in offline training phase is provided, whereby the meta policy can quickly adapt to a new model in online adaptation phase.
\end{itemize}

The rest of this paper is organized as follows. Section 2 presents background and related work. Section 3 outlines the proposed approach and illustrates it through a running example. Section 4 presents the modeling and synthesis process, the meta reinforcement learning for offline training and online adaptation, and the MRL-enabled MAPE-K framework. Section 5 conducts the case study and discusses the experimental results. Section 6 outlines a short conclusion and highlights the future work.

\section{Background and Related Work}
This section presents some background on \emph{meta reinforcement learning} and the related work.

\subsection{Meta Reinforcement Learning}
\emph{Reinforcement learning} (RL) is the study of machine learning methods for learning a policy (i.e., a function mapping states to actions) so as to maximize a numerical reward signal \cite{sutton1998introduction}. 
The basic mathematical model of RL is Markov decision process (MDP), which describes the interaction between an agent and its environment. 
An MDP is defined as a 4-tuple $\mathcal{M}=\left\langle S, A, T, r \right\rangle$, where $S$ is a set of environmental states, $A$ is a set of agent actions, $T:S\times A\times S\rightarrow [0,1]$ is the transition probability function, and $r:S\times A\times S\rightarrow \mathbb{R}$ is the reward function.
When the environmental state cannot be perceived directly by the agent, the partially observable Markov decision process (POMDP) can instead be used as the mathematical model \cite{oliehoek2016aconcise}. 
We compute the optimal policy (i.e., a policy maximizing a numerical reward signal) with dynamic programming, temporal-difference learning or Monte Carlo methods \cite{sutton1998introduction}.

\emph{Meta learning} (ML) is one of the most active fields in machine learning. 
It aims to train a model that quickly adapts to new tasks by using only a small amount of training data and converges in a few training steps \cite{mishra2018simple,finn2017model}. 
A successful ML method is model-agnostic meta-learning (MAML) \cite{finn2017model}, of which the basic idea is to train the learner’s initial parameters on different tasks in advance such that the learner achieves maximal performance on a new task after updating parameters through a few gradient steps.

\emph{Meta reinforcement learning} (MRL) combines reinforcement learning with meta learning, and aims to learn a meta policy that quickly adapts to new tasks \cite{yuxiang2019norml,abhishek2018metareinforcement}. 
The formulation of MRL in this paper is as follows. 
Each task $i$ examined is a variation of MDP, denoted as $\mathcal{M}_i=\{L_{i}, P_i(s^0), T_i(s^{t+1}|s^{t},a^t),H_i\}$, where $L_i$ is a loss function for evaluating the loss of a policy $\pi$, $P_i(s^0)$ is a distribution over the initial state, $T_i(s^{t+1}|s^{t},a^t)$ is a transition probability function, $s^t$ denotes the state at time step $t$, $a^t$ denotes the action taken at time step $t$, and $H_i$ is an episode length.
The meta policy $\pi$ maps a state $s^t$ to an action $a^t$ and minimizes the total losses on different tasks. 
Then, given a distribution $P(\mathcal{M}_i)$ over tasks, the objective of the meta policy (parameterizing with $\theta$) can be described as follows:
\begin{equation}
\begin{split}
 \min_{\theta}&\mathbb{E}_{\mathcal{M}_i\sim P(\mathcal{M})}\Big[
 L_i(\pi(\cdot;\theta'_i))
 \Big ]\\
 \text{s.t.\ } & \theta'_i=\theta-\alpha\nabla_\theta L_i(\pi(\cdot;\theta))
\end{split}
\label{eq:mrl1}
\end{equation}
$\pi(\cdot;\theta)$ can be implemented with a deep neural network, and $\theta$ are the weights of the network.
For instance, there are $n$ MDPs that the MRL algorithm will handle. 
First, Eq.(\ref{eq:mrl1}) is used to compute the optimal meta parameter $\theta$.
Then, when the MRL algorithm needs to adapt to a certain MDP $\mathcal{M}_i$, it iteratively updates the parameter by using $\theta_i=\theta_i-\alpha\nabla_\theta L_i(\pi(\cdot;\theta_i))$ where the initial value of $\theta_i$ is meta parameter $\theta$. 

\subsection{Related Work}
A self-learning adaptive system is able to adjust or re-plan the adaptation policy online. 
Most of the existing approaches attribute the problem to \emph{incomplete information} \cite{nicol2014hope,daniel2020multitier,sykes2013learning}. 
Here, incompleteness lies in different aspects, including the spatial environment, the hardware/software of the targeted system, and the user's objectives. 

From a solution perspective, the relevant studies to ours can be roughly classified into two categories.
The first puts more emphasis on the \emph{model-revision method}, which focuses on reflecting the assumption violation in the domain models. 
Sykes et al. \cite{sykes2013learning} consider that incompleteness lies in the environment domain models constructed for determining the adaptive system behavior.
They propose an approach which incorporates a probabilistic rule learning method in a self-adaptive system to update and correct environment domain models. 
D'Ippolito et al. \cite{nicol2014hope} focus on the uncertainty about the accuracy and completeness of the behavior or architecture models.
They design a framework to support multiple environment models and adapt to the functional behavior of the system through degraded or enhanced operations guided by assumption satisfaction assessment.
Jamshidi et al. \cite{jamshidi2017transfer} train a architecture model (i.e., the configuration-non-functional properties relationship model) using samples from simulator in the offline phase, and transfer it to real environment based on transfer learning.

The second puts more emphasis on the \emph{policy-replanning method}, i.e.  directly adjusting the adaptation policy rather than explicitly repairing the domain models. 
Kim et al. \cite{kim2009reinforcement} use reinforcement learning to enable the evolution of the adaptation policy at run-time. 
Zhao et al. \cite{zhao2017reinforcement} propose to establish a case base (consisting of multiple pairs of model and policy) in the offline phase and retrieve the best-fitting case to serve as a real-taken adaptation policy at run-time. 
By enhancing the work in \cite{nicol2014hope},
Ciolek et al. \cite{daniel2020multitier} further define the multi-tier framework for planning that allows the specification of different sets of assumptions and of different corresponding objectives, and solve problem instances by a succinct compilation to a form of non-deterministic planning.

As far as we know, our study is the first attemp to introduce meta-learning to support a self-adaptive system adjusting its adaptation policy online.

\section{Overview of Our Approach}
This section gives an overview of our approach, and illustrates the main idea through a running example about a robotic system.

\subsection{MeRAP Approach}
Basically, the novel idea underlying MeRAP includes: 
(1) a set of sub-models is pre-defined at development time. Each of the sub-models is about a certain aspect of the environment-system dynamics. The sub-models are often constructed by domain experts in corresponding aspects. These sub-models are synthesized into integrated dynamic models;
(2) a meta reinforcement learning algorithm is used to learn a \emph{meta policy} (which is parameterized) over the multiple models in the offline phase such that the meta policy can quickly adapt to the actual environment-system dynamics in the online phase with a few adaptation operations.

\begin{figure}[htb]
\centering
\includegraphics[width=3.4in]{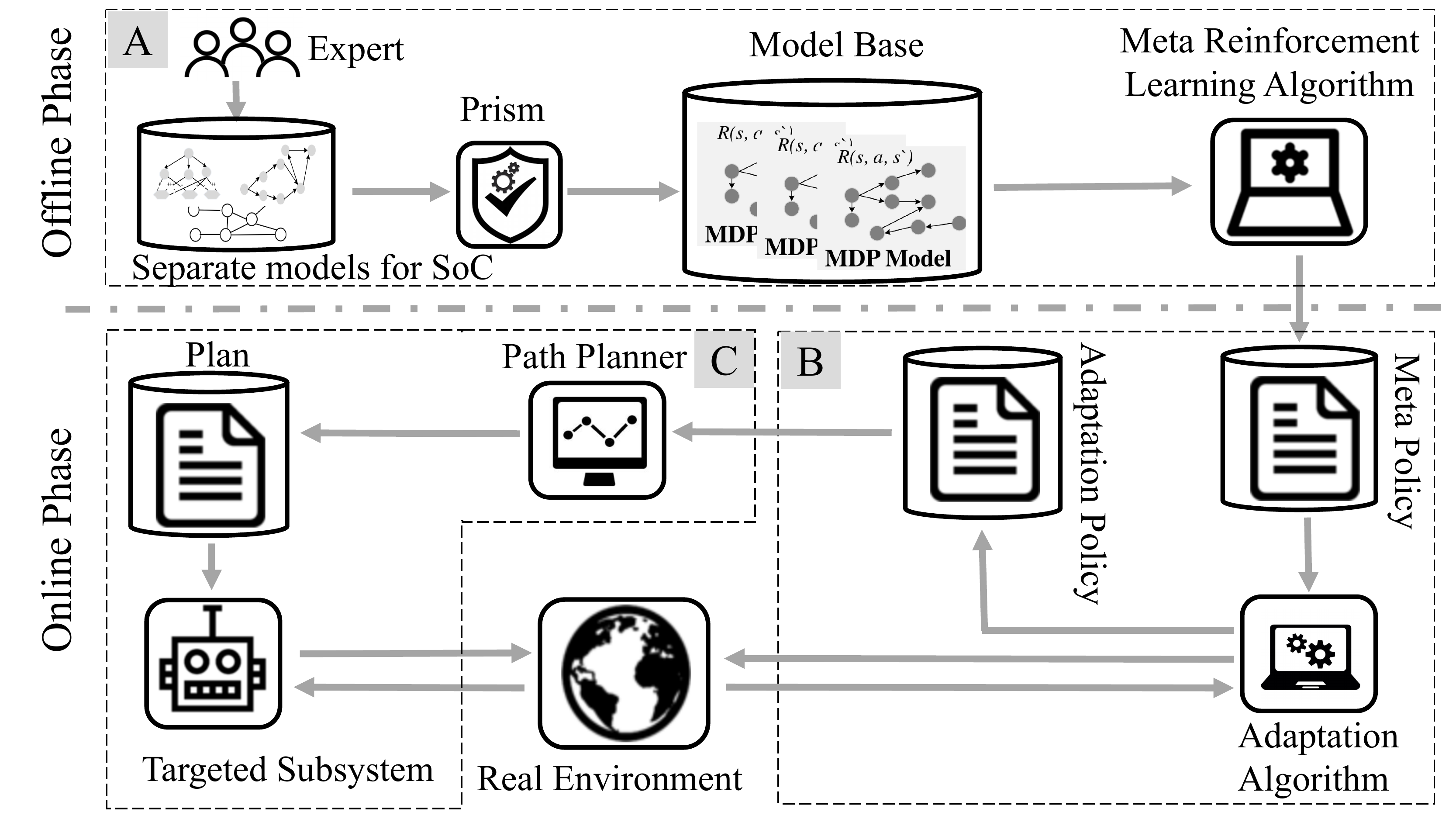}
\vspace{-0.3cm}
\caption {Overview of MeRAP Approach}
\label{fig:pipeline}
\end{figure}

Fig. \ref{fig:pipeline} depicts the overview of MeRAP process. The development cycle can be divided into three phases:

\begin{itemize}
\item \emph{Offline training}: That is for training a meta policy (\textbf{A} in Fig. \ref{fig:pipeline}). 
First, the sub-models on the spatial environment, the system capabilities, and the objectives under different assumptions are defined.
Second, these sub-models are synthesized into integrated dynamic models (represented by MDPs), which constitute a model base. 
Third, a meta reinforcement learning algorithm learns a meta policy over these MDPs in the model base.
 
\item \emph{Online adaptation}: That is for generating a usable adaptation policy (\textbf{B} in Fig. \ref{fig:pipeline}). 
That is the meta policy.
The adaptation algorithm takes the meta policy as its initial policy and interacts with the actual environment. 
Based on the response of the actual environment, the algorithm updates the current policy to produce an adaptation policy suitable to the current environment.
 
\item \emph{Online execution} (\textbf{C} in Fig. \ref{fig:pipeline}). In this phase, the targeted system takes action based on the adaptation policy generated in the online adaptation phase, so as to complete the task of meeting the objective in the actual environment.
\end{itemize}
\vspace{-0.2cm}
\vspace{-0.2cm}

In order to achieve this goal, two novel techniques are adopted in MeRAP:
\begin{itemize}
    \item By following the principle of \emph{separation of concerns} (SoC), we distinguish three aspects that are related to the adaptation policy, i.e., the spatial environment, the system's capabilities, and the users' objectives\cite{tsigkanos2018early,nianyu2020early}. 
    The sub-models with different configurations are designed by domain experts. 
    Then, these sub-models are synthesized to multiple integrated models by using PRISM-games\footnote{A model checker that can support the design of various probabilistic models with multiple models. \url{https://www.prismmodelchecker.org/games/}} \cite{KNPS20}. 
    Let $<\mathcal{E}, \mathcal{C}, \mathcal{O}>$ be the sub-models of the spatial environment, the system capability, and the user objective with a certain configuration respectively each of which is represented by one MDP.
    Assume there are $N$ sub-models for the spatial environment, $M$ sub-models for the system capability, and $K$ sub-models for user objective and they are synthesized into $N\times M\times K$ integrated models.
    \item The \emph{meta reinforcement learning} (MRL) is incorporated into the offline training phase and the online adaptation phase. 
    As well, an MRL algorithm is designed to enable the computation of a meta policy that minimizes the loss function over the integrated models such that the meta policy achieves maximal performance on actual dynamics after updating parameters in a few gradient steps.
\end{itemize}

\begin{table*}[!htbp]
\centering
\algsetup{linenosize=\tiny}
  \caption{Configurations of the running example.}
  \vspace{-0.2cm}
  \label{tab:assumption}
  \begin{tabular}{cccccl}
    \toprule
    & Configuration-1 & Configuration-2 & Configuration-3 & The number of Configurations \\
    \hline
    Spatial environment & B is blocked;   
    & C is blocked;
    & B and C is blocked
    & 3\\
    System capabilities & motor speed is low & motor speed is high & NULL
    & 2\\
    User objectives & G1 is the goal & G2 is the goal & G3 is the goal
    & 3\\
    \bottomrule
  \end{tabular}
  \vspace{-0.3cm}
\end{table*}

\subsection{Running Example}
\begin{figure}[htb]
\vspace{-0.2cm}
\centering
\includegraphics[width=3.4in]{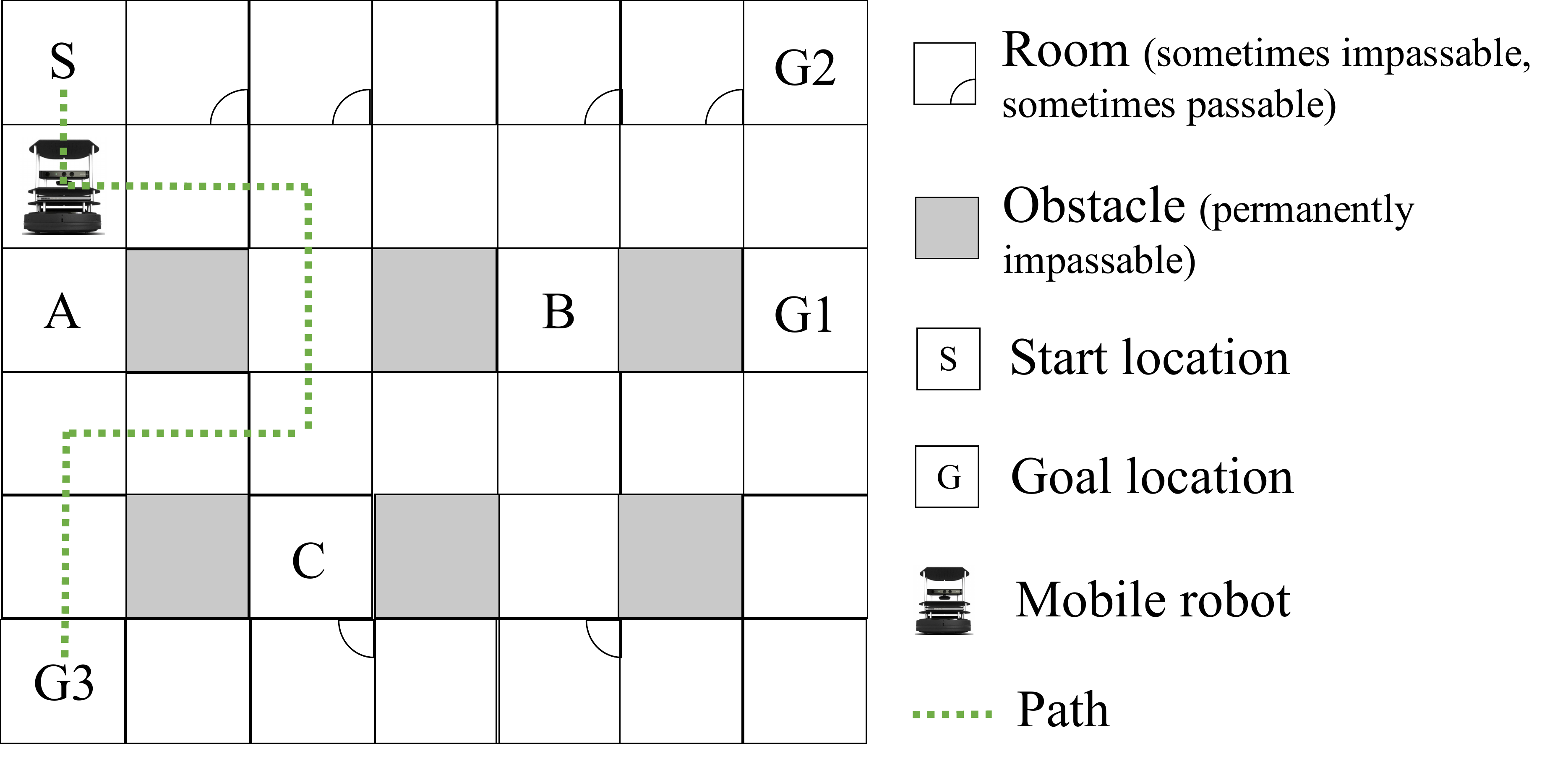}
\vspace{-0.4cm}
\caption{Running example}
\label{exmaple}
\vspace{-0.2cm}
\end{figure}

Consider the robot navigation example shown in Fig.\ref{exmaple}: a mobile robot is ordered to drive from a start location to a goal location in a building. 
The mobile robot is equipped with a laser rangefinder and uses \emph{simultaneous localization and mapping} (SLAM) \cite{simon2020map} to sense the spatial environment. 
Due to the limitations of SLAM, the spatial environment is partially observable to the robot (for example, in Fig.\ref{exmaple}, the mobile robot cannot detect the object behind the wall). 
The accuracy of the robot's sensors and motors may be low or high. 
Because of the different users' objectives, there are various start locations and goal locations.

In such a scenario, the robot as an SLAS needs to cope with the problem of planning with incomplete information about the spatial environment, the system’s capabilities, and the users' objectives. 
In our approach, the following three phases are carried out:

In the offline training phase, 
all possible configurations (shown in Table \ref{tab:assumption}) of the three concerns are identified. 
On the basis of the $3+2+3$ configurations, three spatial environment sub-models, two system capabilities sub-models, and three objective sub-models are independently constructed. 
Next, $3\times 2\times 3$ integrated models represented by MDPs are synthesized by PRISM-games. 
Finally, a meta reinforcement learning algorithm is used to enable the mobile robot to learn a meta-policy over the $3\times 2\times 3$ MDPs.
 
In the online adaptation phase, the meta policy is deployed in the mobile robot. 
The current task assigns $S$ and $G3$ as the start and goal location, respectively; the robot detects that location $A$ is blocked.
With the response of the actual situation, the robot evolves its meta policy to a more suitable adaptation policy.
The green dotted line is the path generated based on the updated adaptation policy.
 
In the online execution phase, the robot takes actions based on the updated adaptation policy and completes the tasks. 
If the current adaptation policy does not perform well, the robot will switch to the online adaptation phase again and a new adaptation policy is generated.

\section{Meta Reinforcement Learning \\ for Online Adaptation}
In this section, we start by describing the modeling and synthesizing method, and detail the key techniques and algorithms used in MeRAP.

\subsection{Modeling and Synthesizing Separate Concerns}
\label{sec:modeling}
For \emph{self-adaptive space systems}, we divide the design concerns  into three aspects, i.e. the \emph{spatial environment}, the \emph{system capability}, and the \emph{user objective}. 
We use location connection graph, probabilistic automaton, and reward function to represent these concerns respectively.

Then, the SLAS is modeled as a 4-tuple $SLAS \triangleq <\mathcal{E},\mathcal{C}, r,\pi>$, where $\mathcal{E}$ is the spatial environment model, $\mathcal{C}$ is the system capability model, $r$ is the objective model represented as a reward function, and $\pi$ is the adaptation policy.

\textbf{Modeling Spatial Environment:} 
The spatial environment model describes the locations,
the attributes of the locations, as well as spatial connectivity. 
The attributes are attached to each location, which may include temperature, light intensity, air humidity, etc., in \emph{self-adaptive space systems}.
Each attribute has a value range and at a certain moment, the attribute has a certain value that can be perceived.

\begin{definition}
\label{def:Space}
The spatial environment model is a 4-tuple $\mathcal{E}\triangleq<P, E, Atr,Det>$, where:
\begin{itemize}
\item $P=\{p_1,...,p_n\}$ is a set of discrete locations;
\item $E=\{e_1,...,e_m\}$ is a set of directed edges. $e_i=[p_s,p_d]$ represents the connection from $p_s$ to $p_d$; 
\item $Atr=\{atr_1,...,atr_k\}$ is a set of attribute value spaces. $atr_i\in Atr$ the value range of $i$th attribute;
\item $Det=\{det_1,...,det_k\}$ is a set of attribute detectors, where $det_i:\rightarrow atr_i$ returns the current value of $i$th attribute
 $\hfill\square$
\end{itemize}
\end{definition}



\textbf{Modeling System Capability:} 
A targeted system (e.g., robot, UAV) is a mobile entity that interacts with the spatial environment, and is controlled by the adaptation policy. The system capability model captures the innate abilities
and the external abilities.

Innate abilities enable the system to take actions that can change the configuration of targeted system itself, e.g., turn off the camera or lower the clock speed of the CPU.
External abilities enable the system to take actions to interact with the spatial environment, e.g., turn left or across a river.
Note that innate abilities can not change the location where the targeted system situates in, while external ones can change the location and even change the spatial environment.

The capability model is defined as a 2-tuple $\mathcal{C}\triangleq (IA, EA)$, where $IA$ is the innate part and $EA$ is the external part.

The innate part is modeled as a probabilistic automaton:
\begin{definition}
\label{def:innate capbility}
The innate capability model is defined as a 4-tuple $IA\triangleq(\mathcal{Q},A_{IA},\delta,F)$, where:
\begin{itemize}
\item $\mathcal{Q}=\{q_0,...,q_n\}$ is a finite set of system states, and $q_0$ is the initial system state;
\item $A_{IA}$ is a finite set of available innate actions;
\item $\delta:\mathcal{Q} \times A_{IA}\times \mathcal{Q} \rightarrow[0,1]$ is the transition probability function of the system;
\item $F$ is a set of terminal system states. 
 $\hfill\square$
\end{itemize}
\end{definition}

$EA$ is formally constructed as a function $EA: P \times A_{EA} \times P \rightarrow [0,1]$, where $P$ is the location set, and $A_{EA}$ is the external action set available to the system. Here, $\forall p\in P, \forall a \in A_{EA}, \sum_{p'\in P}EA(p,a,p')=1$.

\textbf{Modeling User Objective:} 
User's objective is normally a mission that the targeted system needs to achieve
(e.g., search for survivors and transport relief supplies are both tasks). 
The reward function is designed for modeling the achievability of the mission, which assigns a value to each edge for the spatial environment model and the system capability model.
\begin{definition}
\label{def:objective}
The reward function is $r:S \times A \times S \rightarrow \mathbb{R}$, where $S=P\times \mathcal{Q}$, $A=A_{EA}\bigcup A_{IA}$.
 $\hfill\square$
\end{definition}
Here, $S$ is the set of \emph{location-aware system states}, $r(s,a,s')=R$ expresses that the process of taking action $a$ in state $s$ and then jumping to the next state $s'$ has a $R$ reward. 
We assume that the higher the expected accumulated rewards are, the better the goal will be accomplished. 
So the task of finding a policy that leads to optimal results (makes the system reach the goal) can be transformed into an optimization problem, i.e., finding a state-action sequence $\left \langle s_0, a_0,...s_i,a_i,...,s_t, a_t\right \rangle$ that maximizes the expected accumulated rewards\cite{sutton1998introduction}.

\begin{figure}[htb]
\centering
\includegraphics[width=3.4in]{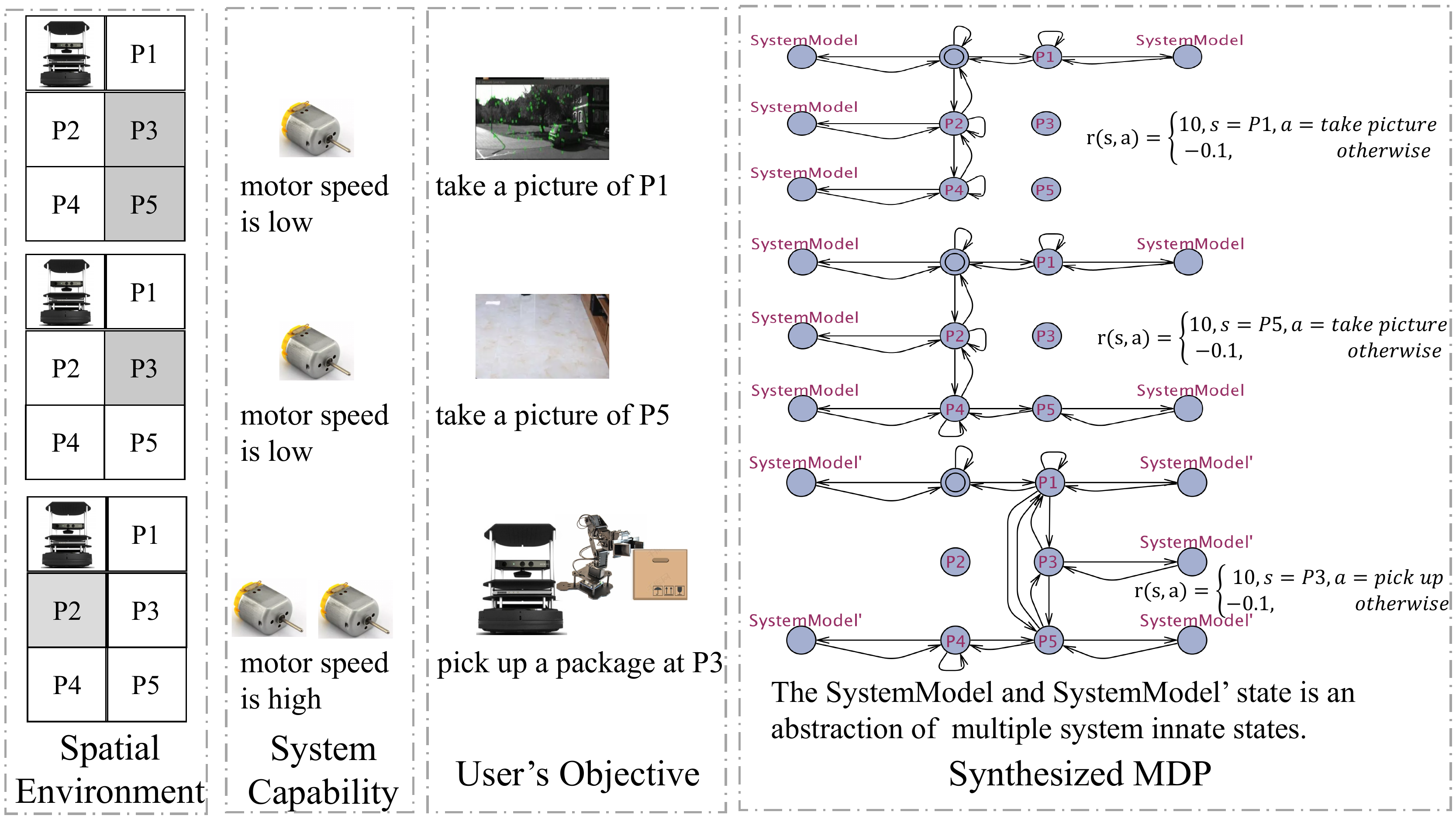}
\vspace{-0.3cm}
\caption{Synthesized Markov Decision Process}
\label{ingegrated}
\vspace{-0.4cm}
\end{figure}

\textbf{Synthesizing Models:} To learn an optimal adaptation policy with the MRL algorithm, the separate sub-models need to be integrated into a \emph{synthesized environment-system model}.
We use an MDP to formally represent the synthesized model.
\begin{definition}
\label{def:mdp}
A synthesized environment-system model is a 4-tuple $\mathcal{M}\triangleq(S,A,T,r)$, where $S=P\times\mathcal{Q}$ is a set of location-aware system states, $A=A_{EA}\bigcup A_{IA}$ is a set of system actions, $T:S\times A\times S\rightarrow [0,1]$ is a transition probability function, and $r:S\times A \times S \rightarrow \mathbb{R}$ is a reward function.
 $\hfill\square$
\end{definition}

The separate sub-models are synthesized into $\mathcal{M}$ as follows:
\begin{enumerate}
\item \emph{Abstracting Separate Models:}
The spatial connectivity are abstracted as a graph $(P,E)$ with semi-automatic or automatic tools (e.g., BIM \cite{eastman2011bim}, CityGML \cite{kolbe2005citygml}). 
And different attributes are assigned to each $p\in P$.
$(P,E)$ and attributes constitute spatial environment sub-model $\mathcal{E}$.
The innate capability $IA$ and the external capability $EA$ are generated from the description of the targeted system.
The user's objective is represented as a reward function $r(s,a,s')$ (the way of generating reward function will be discussed in Section \ref{sec:dis}).

\item \emph{Integrating Spatial Environment Model with External Capability Model:}
For each edge $<p,p'>\in P$, if $\forall a\in A_{EA},EA(p,a,p')\equiv 0$, remove this edge; otherwise, if $<p,p'>$ is not labeled, label the probability $EA(p,a,p')=pr$ and action $a$ on $<p,p'>$, while if $<p,p'>$ is already labeled, add a new multiple edge that is labeled by $pr$ and $a$. 
After these operations, $\mathcal{E}=(P,E,Atr,Det)$ is converted into $\mathcal{E}'=(P, E', Atr , Det)$

\item \emph{Attaching Innate Capability Model:}
For each $p\in P$, add an edge $<p,q_0>$, where $q_0$ is the initial state of $\mathcal{C}$, add $|F|$ edges $<q_F,p>$, where $q_F$ is the terminal state of $\mathcal{C}$. 
Then, integrate the two automatons into a composed automaton $(S,A,T)$, where $S=P\times \mathcal{Q}$, $A=A_{EA}\bigcup A_{IA}$, and $T$ is derived from the $\mathcal{E}'$ and $\delta$.

\item
Assign reward signals to the new automaton by using the reward function $r(s,a,s')$. 
\end{enumerate}

Fig. \ref{ingegrated} illustrates a fragment of the synthesized model.

\textbf{Adaptation Policy:} The adaptation policy is a parameterized function $\pi(a|s;\theta): S\times A\rightarrow [0,1]$, and $\forall s\in S, \sum_{a\in A}\pi(a|s;\theta)=1$, where $S$ and $A$ are defined in \emph{Definition} \ref{def:mdp}. $\pi$ is implemented with deep neural networks (DNN), and $\theta$ are the weights of the DNN.

Given an MDP $\mathcal{M}$, the optimal objective of its corresponding adaptation policy is defined as
\begin{equation}
\setlength\abovedisplayskip{1pt}
\setlength\belowdisplayskip{1pt}
\label{eq:policygoal}
 \pi^o(\cdot;\theta)=\mathop{\arg\max}_{\pi} \mathbb{E}_{s^{t},a^{t}\sim \pi}
 \Big[
 \sum_{t=0}^H \gamma^t r(s^t,a^t,s^{t+1})
 \Big]
\end{equation}
where $\pi^o$ is the optimal policy, $s^t$ and $a^t$ are respectively the MDP state and action at time $t$, $H$ is the terminal time step of $\mathcal{M}$, $T$ is the transition probabilistic function defined in \emph{Definition} \ref{def:mdp}, and $\gamma\in[0,1]$ is the discount rate.

\subsection{Offline Training}
In MeRAP, the SLAS learns a meta policy that works well both on the model base and in the actual environment-system dynamics. 
To this end, we design a meta reinforcement learning algorithm based on MAML \cite{finn2017model}.

The actual environment-system dynamics are modeled as an MDP, denoted as $\mathcal{M}_R$. 
The offline models of the spatial environment, the system capability, and the user objective are denoted as $N_{\mathcal{E}}=\{\mathcal{E}_1,...,\mathcal{E}_n\}$, $N_{\mathcal{C}}=\{\mathcal{C}_1,...,\mathcal{C}_m\}$, and $N_{\mathcal{O}}=\{\mathcal{O}_1,...,\mathcal{O}_k\}$, respectively. $|N|$ is the cardinality of set $N$. 
The model base is a set of offline MDP models $N_{\mathcal{M}}=\{\mathcal{M}_1, ..., \mathcal{M}_{I}\}$, where $I=|N_{\mathcal{E}}|\times |N_{\mathcal{C}}|\times |N_{\mathcal{O}}|$. 
The difference between the offline models and the online actual dynamics is defined as:
\begin{equation}
\setlength\abovedisplayskip{1pt}
\setlength\belowdisplayskip{1pt}
\label{eq:diff}
\begin{split}
\small
 &dif(\mathcal{M}_R,N_{\mathcal{M}})\\
 =&\min_{\mathcal{M}_i\in N_{\mathcal{M}}}
 \Big(
 w_1\sum_{s\in S_i}\sum_{a\in A_i}\sum_{s'\in S_i}(T_R(s,a,s')-T_i(s,a,s'))^2\\
 &+w_2\sum_{s_i\in S}\sum_{a\in A_i}\sum_{s'\in S_i}(r_R(s,a,s')-r_i(s,a,s'))^2
 \Big)
\end{split}
\end{equation}
where $S_i$, $A_i$, $T_i$, and $r_i$ are the elements of $\mathcal{M}_i$, $T_R$ and $r_R$ are the elements of $\mathcal{M}_R$, and $w_1$ and $w_2$ are weight parameters.

For what follows, we impose the following necessary assumption on the problem:
\begin{assumption}
\label{apt1}
The actual environment-system dynamics are distributed on the offline model space, i.e., ${\lim_{|N_{\mathcal{N}}| \to +\infty}} dif(\mathcal{M}_{R},N_\mathcal{M})=0$.
 $\hfill\square$
\end{assumption}
If the problem satisfies \emph{assumption} \ref{apt1}, we can use meta reinforcement learning (MRL) to train a meta policy over the set of MDPs such that the meta policy achieves maximal performance on a new MDP after the parameters of meta policy have been updated using a few adaptation operations.

The loss function of MDP $\mathcal{M}_i$ is defined on the basis of Eq. (\ref{eq:policygoal}):
\begin{equation}
\setlength\abovedisplayskip{1pt}
\setlength\belowdisplayskip{1pt}
 L_i(\pi(\cdot;\theta))=-\mathbb{E}_{s^t,a^t\sim \pi(\cdot;\theta)}\Big[\sum_{t=0}^{H_i}\gamma^t r_i(s^{t},a^t,s^{t+1})
 \Big]
 \label{eq:lossmdp}
\end{equation}
where $H_i$ is the terminal time step of $\mathcal{M}_i$, and $\gamma\in [0,1]$.

Next, using Eq. (\ref{eq:mrl1}) and Eq. (\ref{eq:lossmdp}), the objective of MRL is described as
\begin{equation}
\setlength\abovedisplayskip{1pt}
\setlength\belowdisplayskip{1pt}
\centering
\begin{split}
 \max_{\theta}\sum_{\mathcal{M}_i \sim P(\mathcal{M})}&\Big[
 \mathbb{E}_{s^t,a^t\sim \pi(\cdot;\theta'_i)}[\sum_{t=0}^{H_i}r_i(s^{t},a^t,s^{t+1})]
 \Big]\\
 \text{s.t.\ } &\theta'_i=\theta-\alpha\nabla_{\theta}L_i(\pi(\cdot;\theta))
\end{split}
\label{eq:mrl2}
\end{equation}
where $P(\mathcal{M})$ is the distribution of the MDPs (in this study, we used a uniform distribution), $L_i$ is the loss function of $\mathcal{M}_i$, $\alpha\in(0,1]$ is the learning rate, the policy $\pi(\cdot;\theta)$ is implemented with a DNN, and $\theta$ are the DNN's weights. 
Computing the optimal $\theta$ defined in Eq. (\ref{eq:mrl2}) is called \emph{meta optimization}, which is performed via \emph{stochastic gradient descent} (SGD).

Given $\theta$, $\mathcal{M}_i$ and its loss function $L_i$, $\theta'_i$ is computed as
\begin{equation}
\setlength\abovedisplayskip{1pt}
\setlength\belowdisplayskip{1pt}
\label{eq:adap}
 \theta'_i=\theta-\alpha\nabla_{\theta}L_i(\pi(\cdot;\theta))
\end{equation}
where $\alpha\in[0,1]$ is the learning rate. Then, $\theta$ is computed using $\theta'_i$ computed by Eq. (\ref{eq:adap}):
\begin{equation}
\setlength\abovedisplayskip{1pt}
\setlength\belowdisplayskip{1pt}
\label{eq:meta-opt}
 \theta=\theta-\beta\nabla_{\theta}\sum_{\mathcal{M}_i\sim P(\mathcal{M})}L_i(\pi(\cdot;\theta'_i))
\end{equation}
where $\beta\in[0,1]$ is the meta learning rate. The full meta reinforcement learning algorithm, which stems from that of MAML \cite{finn2017model} for offline training, is outlined in Algorithm \ref{alg:mrl}.

In Algorithm \ref{alg:mrl}, an episode is a state-action-reward sequence from the initial time ($t=0$) to the terminal time ($t=H_i$). 
The inner loop (from line 5 to line 8) computes $\theta_i'$ for MDP $\mathcal{M}_i$ using Eq. (\ref{eq:adap}). 
Line 8 accumulates new episodes for evaluating the gradient in Eq. (\ref{eq:meta-opt}). 
The outer loop (from line 3 to line 10) computes $\theta$ from the inner-loop $\theta_i'$.

\begin{algorithm}[!htb]
\footnotesize{
\caption{MRL Training Procedure}
\label{alg:mrl}
\KwIn{
Model base $N_{\mathcal{M}}$, distribution of MDPs $P(\mathcal{M})$;
adaptation step size $\alpha$, meta learning rate $\beta$, and inner episodes $K$.
}
\KwOut{$\theta$ after trained.}
Initialize $\pi(\cdot;\theta)$ with random $\theta$;\\
\While{not done}
    {   Sample batch of models $\mathcal{M}_{i}$ from $N_{\mathcal{M}}$ with $P(\mathcal{M})$;\\
    \For{\textbf{all} $\mathcal{M}_{i}$}
            {
                Use $\pi(\cdot;\theta)$ to sample $K$ episodes $E=\{<s^0,a^0,r^0,...,s^{H_i}>\}$ in $\mathcal{M}_i$;\\
                Evaluate the gradient using $E$ and $L_i$ in Eq.(\ref{eq:lossmdp});\\
                Update $\theta^{'}_{i}$ using Eq.(\ref{eq:adap});\\
                Use $\pi(\cdot;\theta'_i)$ to sample new $K$ episodes $E'_i=\{<s^0,a^0,r^0,...,s^{H_i}>\}$ in $\mathcal{M}_{i}$;\\
            }
        Update $\theta$ using Eq.(\ref{eq:meta-opt}) and based on data $E'_{i}$;\\
        
    }
}
\end{algorithm}
\vspace{-0.4cm}

\subsection{Online Adaptation and Execution}
In the online adaptation phase, the SLAS uses meta policy $\pi(\cdot;\theta)$ generated by algorithm \ref{alg:mrl} as its initial adaptation policy, senses the actual environment-system dynamics, and adjusts its policy to suit the actual situation.

Particularly, when sensing the actual dynamics, the SLAS accumulates actual environment-system episode data $e_{R}=<s^0,a^0,r^0,...,s^T,a^T,r^T>$, where $T$ is the current time step. 
The adaptation policy is updated with stochastic gradient descent using $e_{R}$:
\begin{equation}
\setlength\abovedisplayskip{1pt}
\setlength\belowdisplayskip{1pt}
\begin{split}
\label{eq:adaptation}
 \theta_R=\theta-\alpha\nabla_{\theta}L_{R}(\pi(\cdot;\theta))
 =\theta +\alpha\nabla_{\theta}\sum_{t=0}^{T}\gamma^t r^t
\end{split}
\end{equation}
where $r^t$ is sampled from $e_R$. To compute the optimal adaptation policy that maximizes the reward in the real environment, we repeatedly solve Eq. (\ref{eq:adaptation}):
$\theta_R=\theta_R +\alpha\nabla_{\theta_R}\sum_{t=0}^{T}\gamma^t r^t$. Each iteration of Eq. (\ref{eq:adaptation}) is called a gradient step.

Finally, in the online execution phase, the targeted system takes actions based on the adaptation policy $\pi(\cdot;\theta_R)$. 
To cope with the changes in the environment, MeRAP also supports an online adaptation-execution switching cycle in which SLAS continually detects the environment-system dynamics. 
The trigger for switching from execution to adaptation is designed as: $\sum_{t=T_1}^{T_2}\gamma^t r^t < TR$, where $TR$ is the threshold reward, $T_1$ is the start time step of monitoring, and $T_2$ is the end time step of monitoring.



\subsection{Integrating MRL into MAPE-K Loop}
\begin{figure}[htbp]
\centering
\includegraphics[width=3.4in]{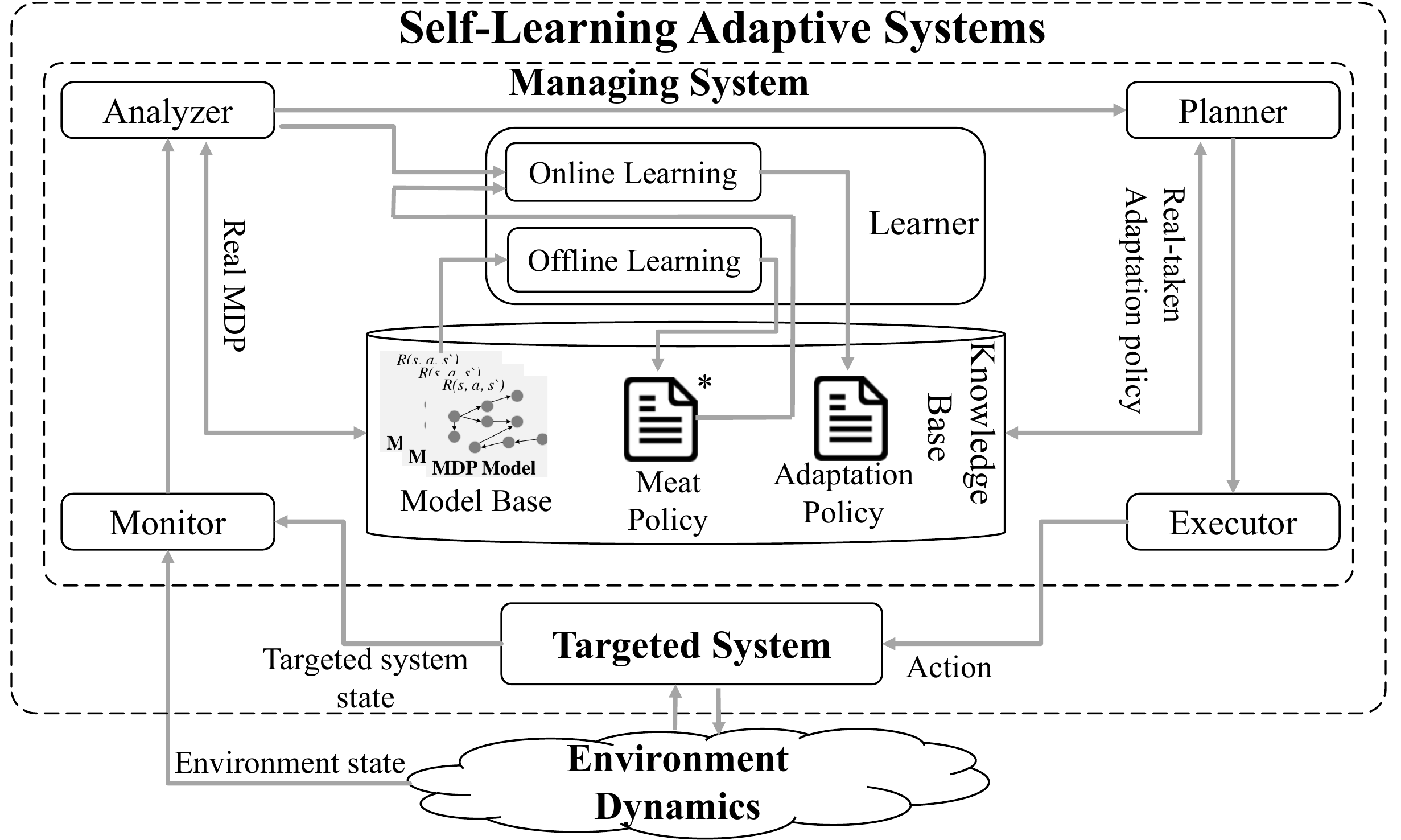}
\vspace{-0.3cm}
\caption{Framework of Self-Learning Adaptive Systems}
\label{Implementation}
\vspace{-0.4cm}
\end{figure}

We propose a new self-adaptive framework, shown in Fig. \ref{Implementation}, incorporating meta reinforcement learning based on the MeRAP approach and MAPE-K loop \cite{kephart2003vision}.

\begin{itemize}
    \item The \textbf{knowledge base} saves and manages the set of MDPs designed in the offline phase, the meta policy generated by the learner, and the adaptation policy adjusted to a real scenario in the online phase. 
    \item The \textbf{learner} works in two phases: in the offline phase, it takes as input all MDPs from the knowledge base and outputs a meta policy for these MDPs by using the MRL algorithm (Algorithm \ref{alg:mrl}). In the online phase, it takes as input the meta policy and actual feedback from the environment and outputs an adaptation policy.
    \item The \textbf{monitor} gathers and synthesizes the dynamics of the targeted system and its environment through sensors.
    \item The \textbf{analyzer} checks whether the targeted system and the environment have changed, generates an MDP for the real domain (environment, targeted system, and objectives), and decides whether the planner needs to re-plan.
    \item The \textbf{planner} generates adaptation actions aiming to counteract the changes in the domain and satisfy the user's goals. The planner uses the real-taken adaptation policy generated by the learner to make an action sequence. 
    \item The \textbf{executor} exerts the adaptation actions on the targeted system.
\end{itemize}

\section{Experiment and Discussion}
This section shows how well MeRAP learns the system's adaptation policy offline and adjusts it online. 
In particular, we conducted case studies on a robotic system (shown in Fig. \ref{exmaple}) and build the MDPs for the system in the way described in Section \ref{sec:modeling}. 
To systematically evaluate the performance of different approaches on the multiple model problem, we considered the dimensions of \emph{coverage} (which refers to whether the cases that happen in the real environment-system dynamics are considered in the offline phase) and \emph{cause} (which refers to whether the concerns, environment, targeted system, and user's objective change online). 
The TABLE \ref{tab:problem} shows the two dimensions.
\footnote{Our project link is: \url{https://github.com/GeorgeDUT/MetaRLSAS}}

The purpose of these experiments is to answer the following research questions:
\begin{itemize}
\item RQ1 (Adaptability): Compared with the baseline approaches, how fast does MeRAP adapt to the real environment-system dynamics or user's objectives, and what is the quality of the policy after adaptation?
\item RQ2 (Cost and quality): What are the effects of MeRAP's parameters on the training time and quality of the adaptation policy? How should parameters be optimized to maximize the system's utility?
\item RQ3 (Online performance): Under the optimal parameters, what is the online performance of MeRAP?
\end{itemize}

The baseline approaches of the evaluation were online policy evolution (OPE) \cite{kim2009reinforcement,zhao2017reinforcement}, which is widely used in evolving the policy with environmental dynamics, and a pre-trained policy, which uses an existing adaptation policy to take actions in the actual environment-system dynamics.

\begin{figure*}[htb]
\centering
\includegraphics[width=7.15in]{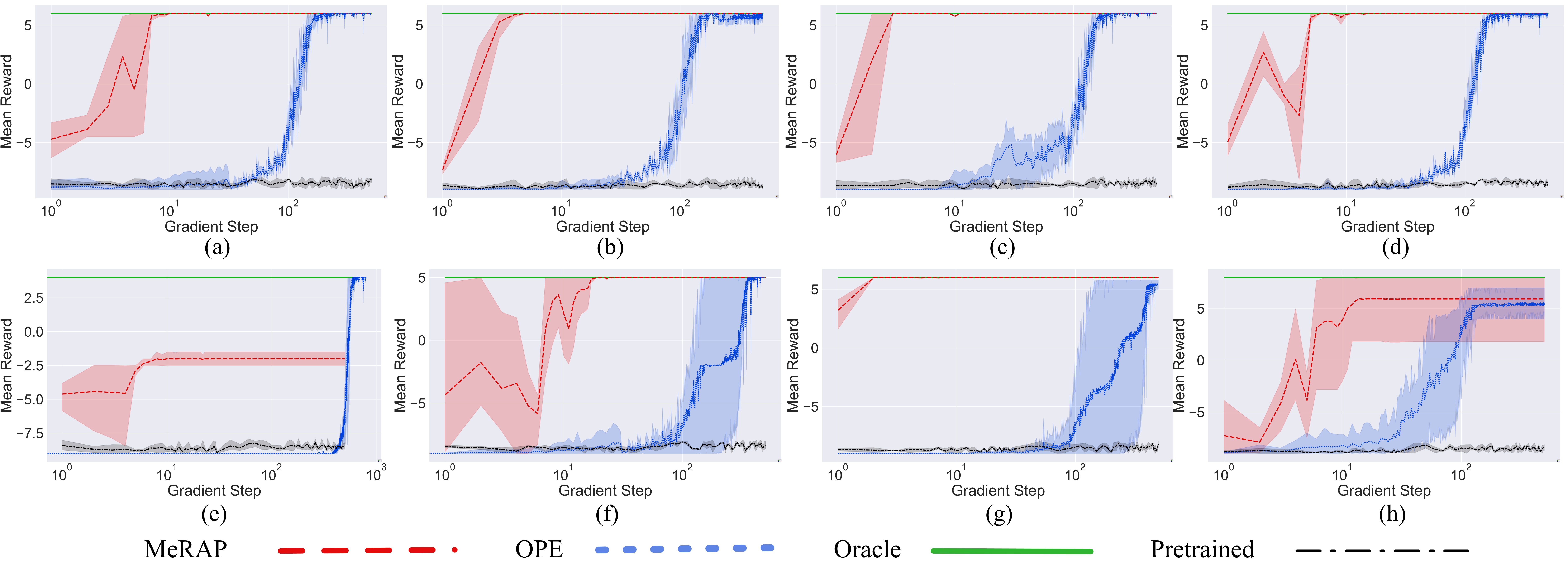}
\vspace{-0.6cm}
\caption{Adaptability results: the \emph{oracle} policy receives the real dynamics and goal position as input and plans the path by using search-based algorithms).}
\label{RQ1}
\vspace{-0.3cm}
\end{figure*}

\subsection{Adaptability}
\label{sec:rq1}
Fig. \ref{RQ1} shows the adaptation curves in different cases. The x-axis represents the gradient step (in each gradient step, the MRL algorithm uses the real episode to update the weights of the neural networks $\theta$), and the y-axis represents the reward (which evaluates the quality of the current policy) after the gradient step is updated. The shadowed area is enclosed by the minimum and maximum values of $15$ experiments, and the bold curve in the middle is the mean. To better show the adaptation process, the horizontal axes are logarithmic scales. Table \ref{tab:problem} shows the correspondence between the adaptation curves and the experimental cases.


\begin{table}[!htbp]
\vspace{-0.3cm}
\centering
\algsetup{linenosize=\tiny}
  \caption{Adaptability for different types of multiple model problem.}
  \vspace{-0.2cm}
  \label{tab:problem}
  \begin{tabular}{cccccl}
    \toprule
    & Objective & Environment & Targeted system & Mixed \\
    \hline
    Covered &Fig.5.(a)&Fig.5.(b) &Fig.5.(c)&Fig.5.(d)\\
    Not covered &Fig.5.(e)&Fig.5.(f)&Fig.5.(g)&Fig.5.(h)\\
    \bottomrule
  \end{tabular}
\end{table}

The results show that (1), in most cases, MeRAP is quicker than the baselines at adapting to the real environment-system dynamics and the user's objectives (no more than ten gradient steps), and it achieves the highest reward; (2) however, when the user's objective is \emph{not covered} (in this case, we set three configurations of the user's objectives, \{G1,G2,G3\} in the offline phase, but the objective is \{G4\} in the online phase), the offline knowledge may mislead MeRAP to a locally optimal solution. We will discuss the second point further in Section \ref{sec:dis}.


\begin{figure*}[htb]
\centering
\includegraphics[width=7.15in]{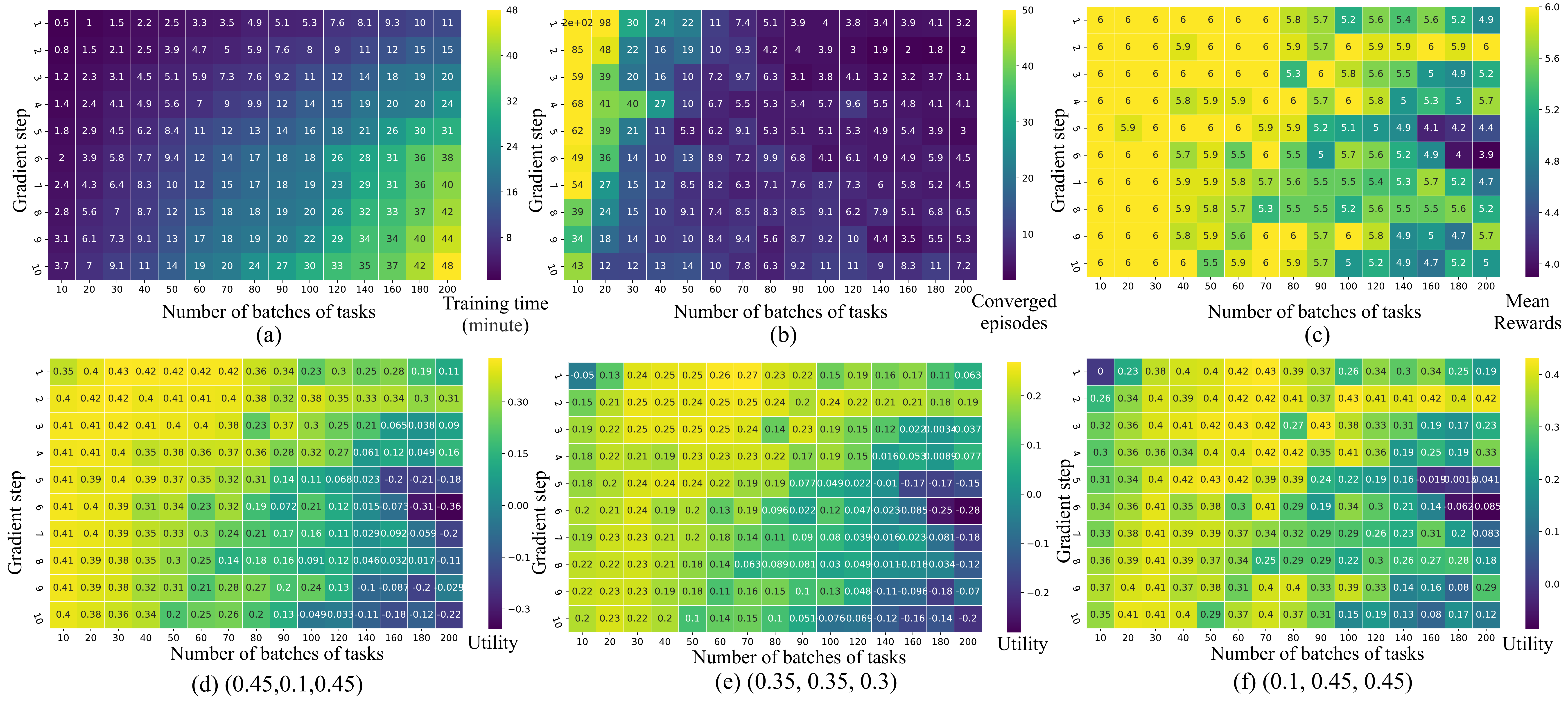}
\vspace{-0.6cm}
\caption{Effect of different parameters on policy quality and offline training time}
\label{RQ2}
\vspace{-0.4cm}
\end{figure*}

\begin{figure}[htb]
\centering
\includegraphics[width=3.3in]{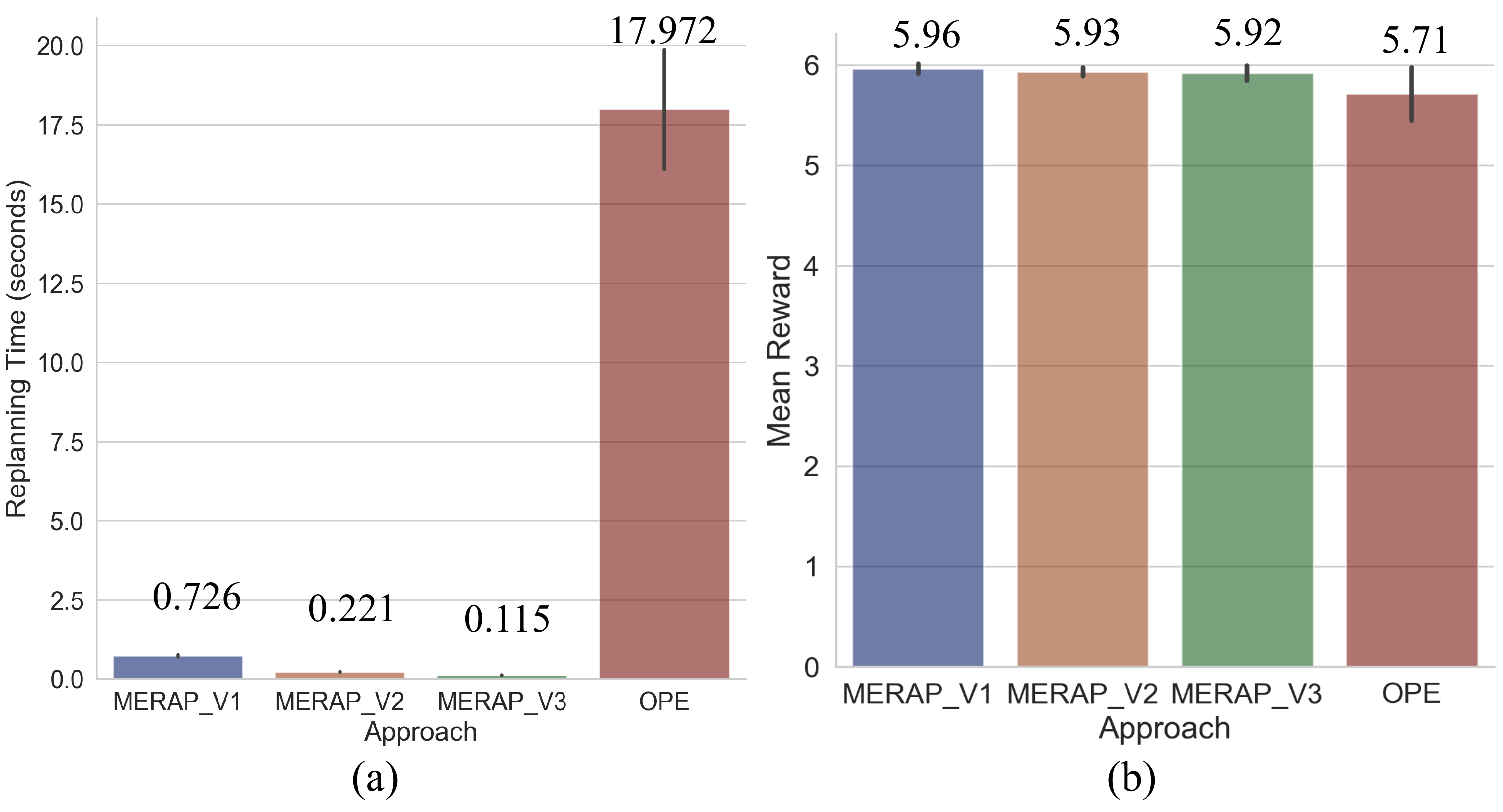}
\vspace{-0.5cm}
\caption{Online re-planning time and policy's quality}
\label{RQ3}
\vspace{-0.4cm}
\end{figure}

\subsection{Cost and Quality}
We tried different parameter combinations to find the optimal parameters for different preferences in policy quality and offline training time. Two kinds of parameter may affect the quality of a meta policy and the meta training time: (1) the batches of tasks, which refers to the number of training trajectories in the offline training phase; (2) the gradient step in the meta training, which refers to the number of gradient updates (Eq. (\ref{eq:meta-opt})). For example, if we use three gradient steps in the meta training, then $\theta$ is computed with: $\theta=\theta-\beta\Big(\nabla_{\theta}\sum_{\mathcal{M}_i\sim P(\mathcal{M})}L_i(\pi(\cdot;\theta'_i))+\nabla_{\theta}\sum_{\mathcal{M}_i\sim P(\mathcal{M})}L_i(\pi(\cdot;\theta'_i))+\nabla_{\theta}\sum_{\mathcal{M}_i\sim P(\mathcal{M})}L_i(\pi(\cdot;\theta'_i))\Big)$.

Fig. \ref{RQ2} (a)-(c) shows the effect of the number of batches of tasks and gradient steps on training time, converged episodes, and mean rewards. The results show that training time, converged episodes, and mean rewards are influenced by both the number of batches and gradient step. Moreover, when the gradient step is fixed, as the number of batches increases, the training time increases, the converged episodes decreases, and mean rewards mostly decrease.

Next, we tried to find a perfect trade-off among the training time, converged episodes, and mean rewards. The values after normalization are denoted as $t$, $e$, and $r$. The utility of the SLAS is
\begin{equation}
\setlength\abovedisplayskip{1pt}
\setlength\belowdisplayskip{1pt}
 u=-\alpha_1 t-\alpha_2 e +\alpha_3 r, \text{\ \ s.t.\ } \alpha_1 + \alpha_2+\alpha_3 = 1
\end{equation}
Given a set of weights ($\alpha_1,\alpha_2,\alpha_3$), we should find an optimal parameter setting that maximizes the utility $u$. We consider three typical preferences: (1) care more about offline time and the online policy's quality ($\alpha_1 =0.45, \alpha_2=0.1,\alpha_3=0.45$); (2) balanced preferences ($\alpha_1 =0.35, \alpha_2=0.35,\alpha_3=0.3$); (3) care more about online adaptation speed and the policy's quality ($\alpha_1 =0.1, \alpha_2=0.45,\alpha_3=0.45$).

Fig. \ref{RQ2} (d)-(f) show the SLAS's utility over two dimensions that correspond to the gradient step and number of batches of tasks, respectively.
The SLAS's utility is maximized when the parameter combination ``(gradient step, number of batches of tasks)'' is set to ``(1,30)'' for preference (1), ``(1,70)'' for preference (2), and ``(3, 90)'' for preference (3). 
The parameter combination in Section \ref{sec:rq1} is ``(1,70)''.

\subsection{Online Performance}
We considered the three parameter configurations above, i.e., [1,30], [1,70], and [3,90] (named MeRAP\_V1, MeRAP\_V2, and MeRAP\_V3), and OPE. The experimental settings were as follows: in the offline phase, the assumptions on the user's objectives were \{G1,G2,G3\}, while in the online phase, the objective was identified as G2. Fig. \ref{RQ3} (a) and (b) show the re-planning time and mean reward of the adaptation policy. MeRAP\_V3 is the quickest to adjust its policy; it takes 0.63\% of OPE's re-planning time. All approaches achieve a nearly optimal reward; MeRAP\_V1's reward is the highest. According to Fig. \ref{RQ1} (a), the order of the training times in the offline phase is MeRAP\_V3 $>$ MeRAP\_V2 $>$ MeRAP\_V1 $>$ OPE (OPE does not need offline training), while the re-planning times have the opposite order: MeRAP\_V3 $<$ MeRAP\_V2 $<$ MeRAP\_V1 $<$ OPE.



\subsection{Discussion}
\label{sec:dis}
MeRAP outperformed the baselines. It appeared to be especially suitable for applications in which the state space and action space are finite and discrete and can be easily modeled as finite MDPs. However, it was problematic in the following regards:

At development time, there are two aspects to be further considered or enhanced.
{\bfseries ``The same kind of tasks''}:
It is difficult to formally define ``the same kind of task''. 
Eq. (\ref{eq:diff}) is an attempt for measuring the difference between two tasks (models). However, this form is so strict that meta reinforcement learning can be applied to many applications which do not satisfy Assumption \ref{apt1}. In the future, we hope to find the capability boundary of meta reinforcement learning algorithms in a more rigorous way.
{\bfseries Molding user's objective}:
Our approach directly models user's objective as reward function.
It also allows us to derive reward from the values that contribute to the goal. Thus, the development of the task model has two steps: (1) Decompose the goal into basic tasks which contribute their own values to the goal. The goal model (such as the $i^*$ SR model \cite{bencomo2013supporting} or the KAOS model \cite{baresi2010adaptive}) can be used here. (2) Assign contribution values as reward signals to edges (actions) in the environment model and the system model. If the relationship between actions and tasks is not clear, the reward signal is set to zero.
In addition, TABLE \ref{tab:assumption} shows a toy example of configurations, while the configurations in our case study are more complex, e.g., the configurations of the spatial environment are all attributes of all positions.

At run time, there are three aspects to be further considered or enhanced.
{\bfseries Safety}: The core methodology of reinforcement learning is \emph{trial-and-error}, i.e., accumulating experience through randomly taking actions to improve the quality of policy, which may lead the self-learning adaptive system to unsafe states \cite{garcia2015comprehensive}. A preliminary study on this problem has been conducted \cite{mallozzi2019aruntime,aria2020learning,sebastian2016safety}, but most of the SLASs developed so far do not have an effective methodology to resolve the problem.
{\bfseries Thrashing}: When a violation (referring to the difference between the offline assumptions and the real environment-system dynamics) is detected, MeRAP goes back to the meta policy and uses it to re-plan the policy. If the trigger condition for switching from the current policy to re-planning it is too sensitive, the SLAS will begin thrashing. To make matters worse, the re-planning may result in the aforementioned unsafe trial-and-error.
{\bfseries Local Optimum}: Fig. \ref{RQ1} (e) shows a disquieting problem that if the offline models do not cover the real user's objectives, MeRAP will be misled by the offline knowledge to a local optimum. The main cause of this problem is that the meta learning algorithms are usually only suitable for similar tasks \cite{finn2017model,abhishek2018metareinforcement,mishra2018simple}. One of the possible solutions is to design as many environment-system dynamics models as possible at development time, which may improve the meta learning algorithm's generalization.

\section{Conclusions}
Integrating machine learning techniques for enhancing the system's self-adaptability is an interesting and promising direction of study. Machine learning techniques can help in different aspects. A usual scenario is implementing machine learning on different components of self-adaptive systems.

This paper focused on improving the self-adaptive system's ability to learn the adaptation policy, the core of the self-adaptive system, in the case of the multiple model problem, i.e., finding an adaptation policy that can tackle different environment-system dynamics models. The purpose is to enable the system to cope with uncertainties by creating policies that respond to changes and make explicit provisions for learning. This paper builds a three-phase policy deployment mechanism, i.e., offline training, online adaptation, and online execution, and further integrates it into a meta-reinforcement-learning adaptive planning approach that learns a meta policy that can cope with multiple Markov decision processes. To give the learning algorithm enough training data, we designed an approach that supports concern separation modeling, automatic model synthesis, and model-based policy learning. We demonstrated its feasibility in a case study.

For the future work, the problem described in Section \ref{sec:dis} remains to be further explored to find better solution. 
To extend the approach to learning the cooperation adaptation policy for collective self-adaptive systems is also a new interesting direction.

\bibliographystyle{IEEEtran}
\bibliography{sample-base}

\vspace{12pt}

\end{document}